\documentclass{ws-ijmpa}

\begin{document}

\markboth{Huan Zhong Huang}{Recent Results on Searches for Pentaquark States from STAR at RHIC}

\catchline{}{}{}{}{}

\title{Recent Results on Searches for Pentaquark States from STAR at RHIC}

\author{Huan Zhong Huang}
\address{Department of Physics and Astronomy, University of California, Los Angeles, CA 90095-1547 \\
	huang@physics.ucla.edu}

\author{For the STAR Collaboration}


\maketitle


\begin{abstract}
We present results on pentaquark searches from nuclear collisions at RHIC with the STAR detector system. An intriguing peak has been observed in the invariant mass distribution of $pK^{+}+\overline{p}K^{-}$ from 18.6 Million d+Au collision events at $\sqrt{s_{_{NN}}} = 200$~GeV. The peak centers at a mass $1528 \pm 2 \pm 5$~MeV/c$^{2}$ and the FWHM $\sim15$ MeV/c$^{2}$ limited by detector responses. The statistical significance of the peak is $4.2\sigma$. Such a state if confirmed is manifestly exotic and implies a family of isospin one states. A weak signal of less statistical significance ($\sim3\sigma$) has been observed in 5.6M Au+Au collision events at 62.4 GeV. Searches in 10.7M Au+Au collision events at 200 GeV yield no significant signal. The Au+Au results neither confirm nor rule out the d+Au observation as a possible state. 
\end{abstract}


\section{Introduction}

The first report of an observation by the LEPS experiment at SPring-8~\cite{Nakano} for a 
narrow $\theta^{+} \rightarrow nK^{+}$ state of mass close to the prediction of a pentaquark state from a 
chiral soliton model~\cite{Diakonov} has stimulated much experimental and theoretical interest. The observed narrow state has been interpreted as a $uudd\overline{s}$ pentaquark exotic state. The narrow state has been subsequently observed by several other experiments in $nK^{+}$ or $pK_{S}$ 
channels~\cite{DIANA,CLAS1,SAPHIR,CLAS2,COSY-TOF,ZEUS}. The measured width is believed to be very narrow, mostly limited by detector resolution. The analysis of the $KN$ scattering data also indicated that the width must be very narrow, less than a few MeV/c$^2$, if $\theta^{+}$ exists as a resonance state~\cite{arndt,nussinov}.
However, many experiments~\cite{Babar,BES,CDF,HERA-B} reported null results. Recently a CLAS experiment~\cite{clas-2} has indicated that a previously reported pentaquark observation~\cite{SAPHIR} cannot be confirmed with a high statistics measurement. There is no definitive experimental observation of the pentaquark state that is beyond any doubt arising from limited statistics and possible instrumental or kinematic biases~\cite{dzierba}. On the other hand, we do not know the dynamics and the cross section for production of any pentaquark state. It is possible that the production is very small and is favored in nuclear reactions over that in e$^{+}$e$^{-}$ collisions.

Theoretically several models including chiral soliton model~\cite{Diakonov}, diquark~\cite{Jaffe},quark cluster~\cite{Lipkin},
QCD Sum Rule~\cite{Zhu} and others~\cite{Shuryak,Stancu,Hosaka} have been proposed to explain the pentaquark structure
of the observed state. A recent review of theoretical models can be found in reference~\cite{Zhu-1}. Though most models favor a pentaquark state of isospin 0, other isospin configurations of pentaquark states have also been 
proposed~\cite{capstick,ma}. 

We will report on STAR result of searches for a pentaquark state $\theta^{++}\rightarrow p + K^{+}$ from d+Au, Au+Au and Cu+Cu collisions. This state would be an isospin partner of the $\theta^{+}$ if the pentaquark state has non-zero isospin. The double-strange pentaquark state $\Xi^{--}\rightarrow \Xi^{-}+\pi^{-}$ and the decay channel 
$\theta^{+}\rightarrow p + K_{S}$ have also been searched in STAR. We have found no significant signal in these channels from our data samples. We will focus on the $\theta^{++}$ searches in this report.

STAR is one of the large detector systems at Relativistic Heavy Ion Collider (RHIC) at Brookhaven National Laboratory (BNL). The central tracking device is a large solenoidal Time-Projection Chamber (TPC), approximately 4 meters in diameter and 4 meters in length, inside a solenoid magnet. The TPC provides excellent momentum resolution for charged particles and measures the specific ionization energy loss (dE/dx) in the TPC gas. The dE/dx measurement has been used to identify charged particles in nuclear collisions at RHIC: $\pi$-kaon separation up to 0.6 GeV/c and $\pi$-proton separation up to 1.0 GeV/c. Details of the STAR detector system and the TPC can be found in reference~\cite{star}.
    
\section{Results from d+Au Collisions}

We have analyzed 18.6 M minimum bias d+Au collision events at $\sqrt{s_{_{NN}}}=200$ GeV. Note the average number of binary nucleon-nucleon collisions is approximately 7.5 for d+Au collisions. Therefore, the equivalent number of p+p collisions for the d+Au dataset is near 140 M events. The STAR detector system has a $2\pi$ azimuthal coverage and an approximate pseudo-rapidity coverage of $|\eta|<1.5$. Comparing to many other experiments both the amount of our data volume and the large acceptance of our detector system are very unique. Our data, which have been taken with a minimum bias trigger, are dominated by soft particle production in the low transverse momentum ($p_T$) region. 

\begin{figure}
\begin{center}
  \epsfig{file=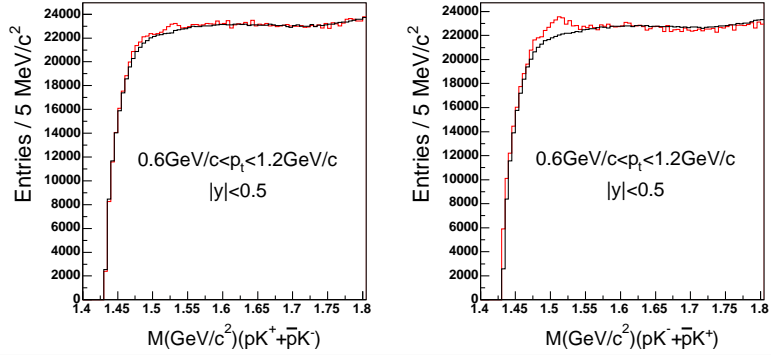,angle=0,height=5cm}
  \caption{Invariant mass distributions of $pK^{+}+\overline{p}K^{-}$ (left) and $pK^{-}+\overline{p}K^{+}$ (right) from d+Au collisions at 200 GeV. The histograms are combinatorial backgrounds from event-mixing technique.}
\label{raw}
\end{center}
\end{figure}

\begin{figure}
\begin{center}
  \epsfig{file=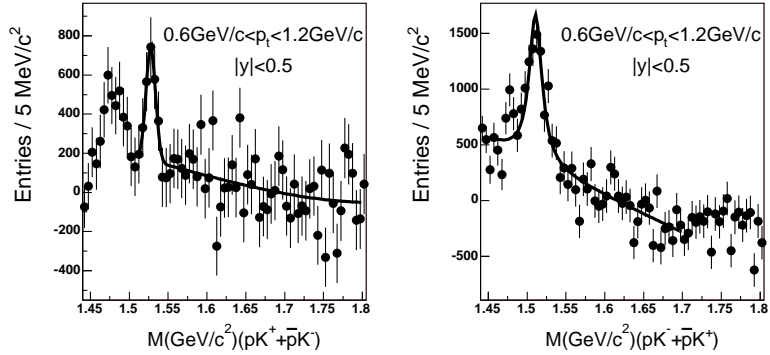,angle=0,height=5cm}
  \caption{Invariant mass distributions of $pK^{+}+\overline{p}K^{-}$ (left) and $pK^{-}+\overline{p}K^{+}$ (right) from d+Au collisions at 200 GeV after the subtraction of combinatorial background. The curves are a fit of Gaussian and polynomial function.}
\label{subt}
\end{center}
\end{figure}

Figure~\ref{raw} shows the invariant mass distributions of $pK^{+}+\overline{p}K^{-}$ (left) and $pK^{-}+\overline{p}K^{+}$ (right) from d+Au collisions at 200 GeV. We select kaon and proton candidate tracks by requiring their dE/dx measurements to be within $2\sigma$ of their respective kaon and proton ionization energy loss curve, where $\sigma$ is the dE/dx resolution depending on the number of fit points and the angle of the track. We have used nn upper momentum cut of $0.6$ GeV/c for kaons and $1.5$ GeV/c for protons in order to select the range where dE/dx particle identification is effective for $\pi$-kaon and $\pi$-proton separation. These track requirements are commonly used in analyses of STAR data~\cite{phi}. 

The histograms in Figure~\ref{raw} are combinatorial backgrounds which were obtained from the event-mixing technique. We choose two events with the characteristics of similar multiplicity and nearby primary vertex location in the TPC. The mixed spectra were obtained by combining kaons from one event and protons from the other event. This event-mixing technique has been established  for analysis of resonance production in nuclear reactions~\cite{mix} and has been used for $\phi$~\cite{phi}, $\rho$~\cite{rho} and $K^{*}$~\cite{kstar} analyses in STAR.

Figure~\ref{subt} shows the combinatorial background subtracted invariant mass distributions. A prominent $\Lambda(1520)+\overline{\Lambda}(1520)$ peak is observed in the $pK^{-}+\overline{p}K^{+}$ invariant mass spectrum (right). A narrow peak centering at $1528\pm 2$ MeV/c$^2$ is observed in the $pK^{+}+\overline{p}K^{-}$ invariant mass spectrum. The width of the peak is approximately 15 MeV/c$^2$ (FWHM), dominated by the detector resolution. The systematic uncertainty on the mass determination is estimated to be 5 MeV/c$^2$ largely due to energy loss and angular uncertainties in track reconstruction. The statistical significance of the peak is approximately $4.2\sigma$. We shall refer this peak as the $\theta^{++}$ for notation purpose only.

The broad mass peak in the low mass region around 1460 MeV/c$^2$ is due to $\Delta^{++}\rightarrow p+\pi^{+}$ 
and $\Delta^{0}\rightarrow p+\pi^{-}$, respectively, in $\theta^{++}$ and $\Lambda(1520)$ spectra, where the pion has been misidentified as a kaon. We have taken dE/dx identified pions, and assign these pions as kaons to combine with protons for invariant mass calculation. We have observed a broad peak with similar features as the residual background in the $\theta^{++}$ and the $\Lambda(1520)$ mass distribution. By rising the momentum upper cutoff above 0.6 GeV/c we can increase the amount of pion contamination in the kaon sample. We found that the amount of residual background increases quickly with the pion contaminations. However, even without an upper momentum limit on the kaon sample the $\theta^{++}$ peak persists, but the statistical significance is reduced considerably to slightly above $3\sigma$ due to a significant increase in the background.  

We have estimated the $\theta^{++}$ yield to be small. Our estimated $\theta^{++}$ to $\Lambda$ ratio is on the order of 
$0.4\%$, below the $95\%$ confidence level upper limit of $0.92\%$ from HERA-B experiment~\cite{HERA-B}. Because there is no effective way to reduce the combinatorial background in the $\theta^{++}$ analysis, it is indeed a challenge for an experiment to achieve the statistical sensitivity necessary to confirm our observation. 

\section{Results from Au+Au Collisions}

We have analyzed Au+Au collisions at RHIC, 5.6 M events at 62.4 GeV and 10.7 M events at 200 GeV. The track selection requirements are the same as that in d+Au analysis. We have only used events with $20$-$80\%$ collision centrality in our pentaquark analyses. For the most central $20\%$ collision events the combinatorial background rises much more rapidly than the particle yield. For analyses of $\phi$ meson production we have already found that it is more difficult to extract $\phi$ yields from the most central Au+Au collisions than from peripheral collisions~\cite{phi}. The $\phi$ yield is a few order of magnitude higher than the estimated yield of $\theta^{++}$ discussed here.

\begin{figure}
\begin{center}
  \epsfig{file=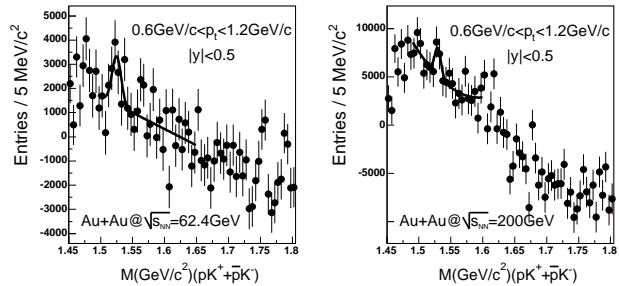,angle=0,height=4.cm}
  \caption{Invariant mass distributions of $pK^{+}+\overline{p}K^{-}$ from Au+Au collisions at 62.4 GeV (left) and at 200 GeV (right) after subtraction of combinatorial background. The curves are a fit of Gaussian and polynomial function.}
\label{au-au}
\end{center}
\end{figure}

Figure~\ref{au-au} shows the invariant mass distributions of $pK^{+}+\overline{p}K^{-}$ from Au+Au collisions at 62.4 GeV (left) and at 200 GeV (right). The combinatorial backgrounds from event-mixing have been subtracted. There is no statistically significant $\theta^{++}$ signal observed in these collision samples: there may be a weak signal of $3\sigma$ in the 62.4 GeV data and of $2\sigma$ in the 200 GeV data. The residual background from mis-identified pions in $\Delta^{++}$ decays appears to be significant.

We have also studied 16.5 M Cu+Cu collision events at 62.4 GeV. No signal has been observed in this dataset.

\section{Discussion}

We have observed an intriguing signal from $pK^{+}+\overline{p}K^{-}$ invariant mass distributions in d+Au collisions at 200 GeV. Although the statistical significance is only $4.2\sigma$, the signal seems to persist regardless many tests in the analyses that were made to increase the amount of pion contamination in the kaon candidate sample. However, searches in Au+Au and Cu+Cu collisions failed to yield a definitive confirmation of the observed peak. 

\begin{figure}
\begin{center}
  \epsfig{file=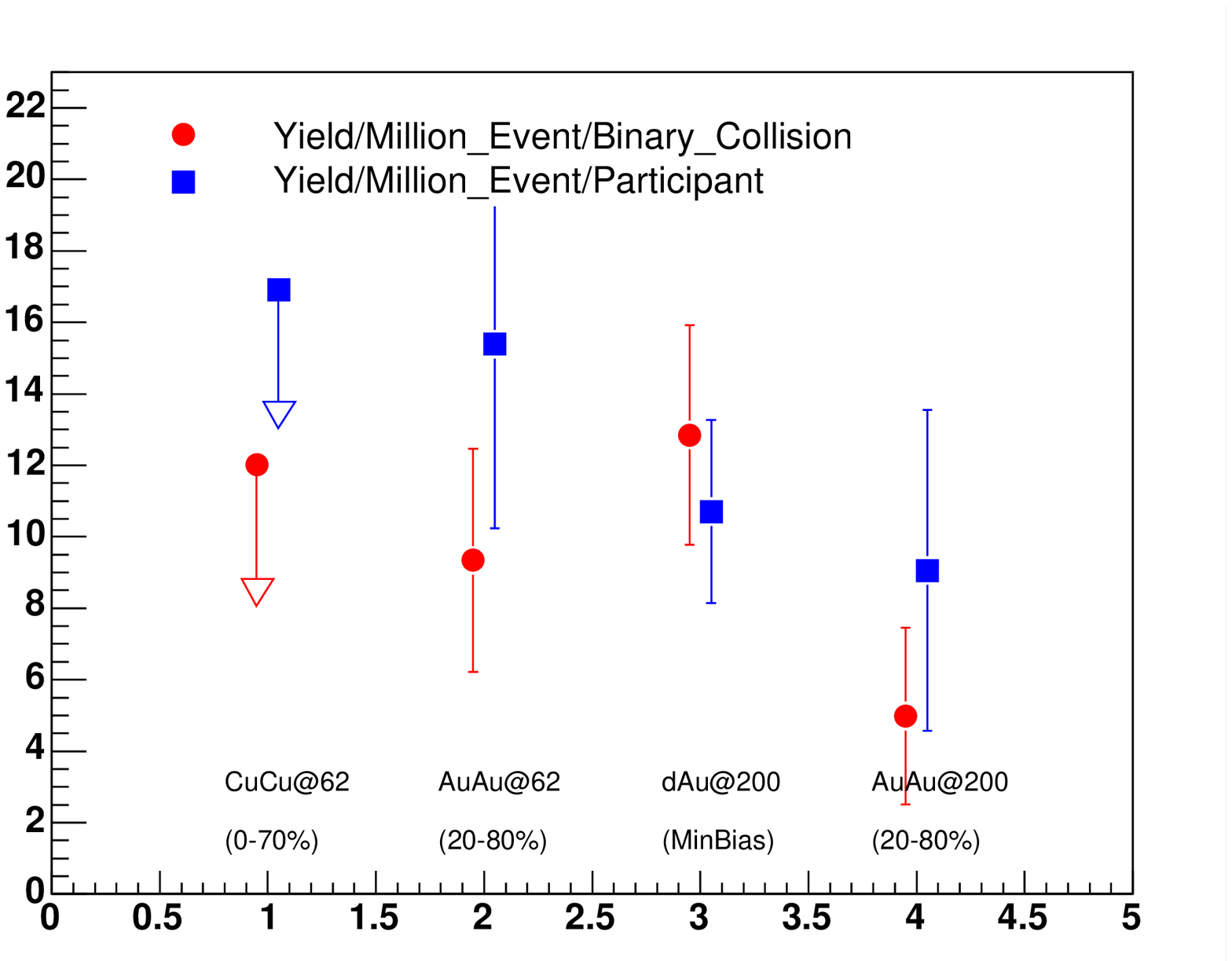,angle=0,height=4.cm}
  \caption{Yield of $\theta^{++}$ per million events normalized by the number of participants (binary collisions) for d+Au and Au+Au collisions. The data for Cu+Cu is a $90\%$ confidence level upper limit.}
\label{all}
\end{center}
\end{figure}

We have studied whether the null results from the Au+Au and Cu+Cu collisions present a bound on the $\theta^{++}$ production that would contradict the observation in d+Au collisions. Figure~\ref{all} shows the observed yield per million events normalized by either the number of participants or the number of binary nucleon-nucleon collisions. Note the possible signals in Au+Au collisions are not statistically significant as indicated by the large statistical errors and their consistency with zero within a few sigma. The limit from Cu+Cu is at the $90\%$ confidence level. There is no obvious contradiction in the measured level of sensitivity from the colliding systems which we have studied. We cannot confirm the d+Au result from the Au+Au and Cu+Cu collisions. Neither can we rule out the possibility that these null results are due to the overwhelming combinatorial backgrounds.

The d+Au, or p+A, collisions could play a special role in searches of pentaquark states in nuclear collisions. Particle yield in p+A collisions is typically higher than that in p+p while the combinatorial background is not overwhelming yet. Recently S. Sarkar et al.~\cite{oset} pointed out that the $\Delta$ and $K^{+}$ interaction is attractive in the vector meson exchange channel in contrast to the repulsive interaction between a nucleon and $K^{+}$. It is possible that a pentaquark state can be formed through $\Delta$-$K^{+}$ coalescence in the final state~\cite{zhang}. In such a scenario nuclear collisions would be more favored for the production of pentaquark states than $e^{+}e^{-}$ collisions and photo-production. STAR will argue for another d+Au run at RHIC with pentaquark searches as one of the important physics objectives. An enhanced data-taking rate and new Time-of-Flight upgrade of the STAR detector system will enable us to seek a definitive answer from a future long d+Au run at RHIC. 

\section*{Acknowledgements}

Huan Zhong Huang wishes to thank Prof. Zongye Zhang from IHEP (Beijing), Prof. Bo-qiang Ma from Beijing University, Dr. An Tai and Jingguo Ma from UCLA for many enlightening discussions. We thank our funding agencies for financial support and the RHIC operations group for technical support.

\end{document}